\documentclass{ws-procs9x6}
\usepackage[absolute]{textpos}

\begin{document}

\title{BURSTING AND SYNCHRONY IN NETWORKS OF MODEL NEURONS}

\author{C.~Geier\textsuperscript{1,2,*}, A.~Rothkegel\textsuperscript{1,2,3}, K.~Lehnertz\textsuperscript{1,2,3}}

\address{
    \textsuperscript{1}Department of Epileptology, University of Bonn,\\ Sigmund-Freud-Stra{\ss}e~25, 53105~Bonn, Germany\\
    \textsuperscript{2}Helmholtz-Institute for Radiation and Nuclear Physics, University of Bonn,\\ Nussallee~14--16, 53115~Bonn, Germany\\
    \textsuperscript{3}Interdisciplinary Center for Complex Systems, University of Bonn,\\ Br{\"u}hler Stra{\ss}e~7, 53175~Bonn, Germany\\
    \textsuperscript{*}E-mail: \texttt{geier@uni-bonn.de}
        }

\begin{abstract} Bursting neurons are considered to be a potential cause of
over-excitability and seizure susceptibility. The functional influence of these
neurons in extended epileptic networks is still poorly understood. There is
mounting evidence that the dynamics of neuronal networks is influenced not only
by neuronal and synaptic properties but also by network topology. We
investigate numerically the influence of different neuron dynamics on global synchrony in neuronal networks with complex connection topologies.
\end{abstract}

\begin{textblock*}{14cm}(3cm,24cm)
\noindent R. Tetzlaff and C. E. Elger and K. Lehnertz (2013), \emph{Recent Advances in Predicting and Preventing Epileptic Seizures}, page 108--116, Singapore, World Scientific.\\
    Copyright 2013 by World Scientific.
\end{textblock*}
\bodymatter

\section{Introduction} 
Epilepsy is a disorder of the brain characterized by an enduring predisposition to generate epileptic seizures and by the neurobiologic, cognitive, psychological, and social consequences of this condition \cite{Fisher2005}. Approximately 1\% of the world's population suffers from epilepsy. An epileptic seizure is defined as a transient occurrence of signs and/or symptoms due to abnormal excessive or synchronous neuronal activity in the brain \cite{Fisher2005,Engel2006}. In about 25\% of individuals with epilepsy, seizures cannot be controlled by any available therapy. During the last decades a variety of potential seizure-generating (ictogenic) mechanisms have been identified, including synaptic, cellular, and structural plasticity as well as changes in the extracellular milieu. Although there is a considerable bulk of literature on this topic (see Ref. \refcite{Engel2007} for a comprehensive overview) the exact mechanisms are not yet fully explored. On the cellular level, bursting neurons are considered to be a potential cause of over-excitability and seizure susceptibility. In regular-firing neurons, a brief depolarization causes the generation of a single action potential, whereas a prolonged depolarization induces a series of independent action potentials. In bursting neurons, threshold depolarization triggers a high-frequency, all-or-none burst of action potentials \cite{Lisman1997,Beck2008}. Although a high abundance (up to 90\%) of bursting neurons can be observed in epileptic tissue \cite{Yaari2002} the functional impact of these neurons in extended epileptic networks is still poorly understood.

Over the past few years, substantial progress has been made in modeling epileptic phenomena at different scales (see Refs. \refcite{Lytton2008} and \refcite{Soltesz2008} for an overview).
Large-scale network models take into account intrinsic properties of neurons and the complex, nonrandom connectivity of cortex \cite{Netoff2004,Percha2005,Dyhrfjeld-Johnsen2007,Feldt2007,Morgan2008,Bogaard2009,Rothkegel2011}.
Findings obtained with these models stress the importance of both cellular and network mechanisms in the generation of seizure-like dynamics, which suggests that a single ictogenic mechanism alone may not be responsible for seizure generation.

We here investigate numerically the influence of different neuron dynamics (regular spiking, chattering, bursting) on global synchrony in neuronal networks with connection topologies of lattice, small-world, and random type.

\begin{table}
\tbl{Model parameter settings for different neuron dynamics.}
{\begin{tabular}{ |l|cccc| }
\hline
neuron dynamics & $a$ & $b$ & $c$ & $d$ \\
\hline
spiking & $0.02$ & $0.2$ & $-65$ & $8$ \\
bursting  & $0.02$ & $0.2$ & $-55$ & $4$ \\
chattering  & $0.02$ & $0.2$ & $-50$ & $2$ \\
\hline
\end{tabular}}
\label{cg-tab-neuronparams}
\end{table}

\begin{figure}[!h] \centering
        \includegraphics[ width = 1.0 \columnwidth ]{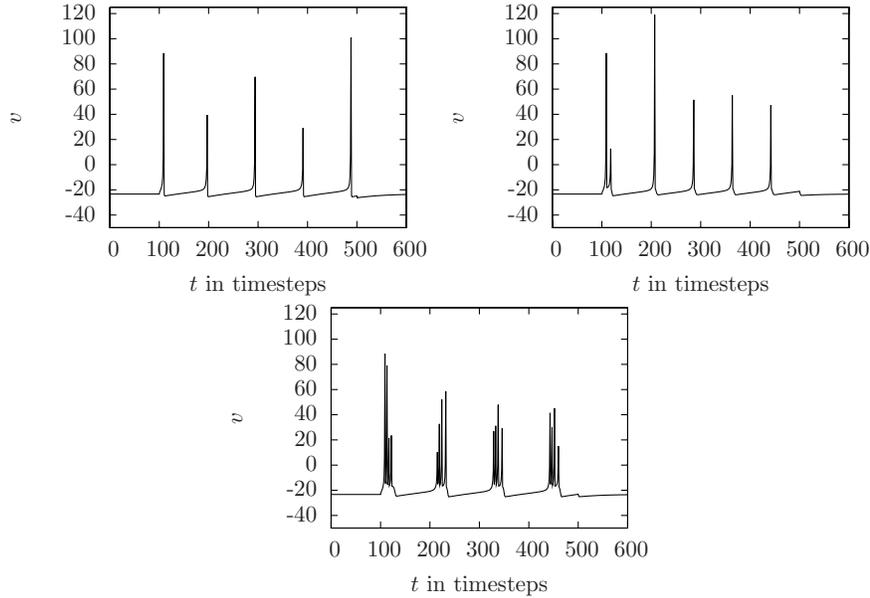}%
        \caption{Temporal evolution of the membrane potential $v$ of different model neurons. Top-left: regular
spiking neuron; top-right: bursting neuron; bottom: chattering neuron.}
        \label{cg-neuronTypesBilder}
\end{figure}

\section{Methods} 
We study networks of Izhikevich model neurons $n$
which are described by the following two-dimensional map \cite{Izhikevich2003a}:

\begin{eqnarray}
v_n(t+1) &=& 0.04 v_n(t)^2 + 5 v_n(t) + 140 - u_n(t) + I(t) \nonumber \\
u_n(t+1) &=& a_n(b_n v_n(t) -u_n(t)).
\label{cg-eq-izhi}
\end{eqnarray}

Here $v_n$ is considered as membrane potential and $I(t)$ specifies the total input current which is composed of three currents: 
\[
 I(t) = I_{\rm{const}} + I_{\rm{noise}  } (t)+ I_{\rm{c} } (t).
\]
$I_{\rm{const }} = 4$ is a constant current which is injected into all neurons. $I_{\rm{c}}(t) := \epsilon \cdot (\mbox{\# presynaptic neurons firing at } t - 1)  $ represents the synaptic coupling, and the noisy current $I_{\rm{noise}}(t)$ is used to generate asynchronous states as initial conditions.

Whenever $v_n(t)$ reaches a threshold (here $30$), neuron $n$ fires and its dynamical variables are updated in the following way:
\begin{eqnarray*}
\text{if } v_n(t) \geq 30 \text{ , then }
\begin{cases} v_n(t) := c_n  &
\\
u_n(t)  := u_n(t) + d_n & 

\end{cases}\\
\end{eqnarray*}

Depending on parameters $a_n,b_n,c_n$, and $d_n$, the neuron model mimics the behavior of regular spiking, bursting, or chattering neurons (cf. Table \ref{cg-tab-neuronparams} and Figure \ref{cg-neuronTypesBilder}).

We studied networks of 10.000 model neurons (we note that we obtained similar findings for networks of 4.000, 20.000 and 50.000 neurons). We considered a one-dimensional lattice on which every neuron is connected to its $k$ nearest neighbors (here $k=20$) using a cyclic boundary condition. Starting from this configuration, every directed connection was removed with probability $\rho \in[0,1]$ and a connection between two randomly chosen, previously unconnected neurons was introduced.

Besides networks consisting of only a single neuron type, we built inhomogeneous networks consisting of spiking and bursting neurons and of spiking and chattering neurons and investigated the network dynamics in dependence on the fraction of different neuron types.

As initial condition we chose asynchronous states, which were generated in the following way: To every neuron we assigned a binary noise input, which takes a value of $I_{\rm{noise }}$ with a probability of $0.1$ and 0 otherwise. We began with $I_{\rm{noise} } = 40$ and repeatedly decreased $I_{ \rm{noise} }$ by $1$ after 200 timesteps until $I_{ \rm{noise} }=0$. Then we let these networks evolve for 10.000 time steps to ensure that transients died out. 
We then measured the fraction of firing neurons $f(t)$ for 2.000 time steps. During this time frame $f(t)$ typically exhibited regular oscillations (alternating periods of high and low values of $f(t)$). 
To assess synchrony in the network, we used the number of concurrently firing neurons at the periods of high values. This number can be estimated (due to the regularity) by the maximum value of $f(t)$ within the observation time $F=\max\{ f(t) | 0 < t < 2.000  \}$. 

We investigated synchrony $F$ depending on the rewiring probability $\rho$ and on the coupling strength $\epsilon$ for both homogeneous and inhomogeneous networks.

\section{Results} 
\subsection{Homogeneous Networks}
\label{cg-Homo}

\begin{figure}[!h] \centering
    \includegraphics[ width = 0.9 \columnwidth ]{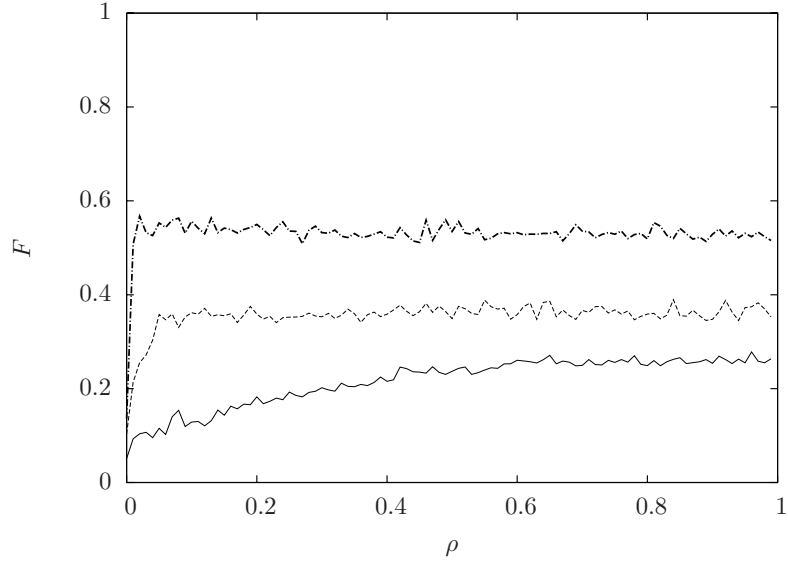}
	\caption{Synchrony $F$ in dependence on the rewiring probability $\rho$; fixed coupling strength $\epsilon = 3.0$. Solid line: spiking neuron network; darker dashed line: bursting neuron network; lighter dashed line: chattering neuron network.}
        \label{cg-hom_cut_rho}
\end{figure}

In Figure \ref{cg-hom_cut_rho} we show the dependence of synchrony $F$ in homogeneous networks on the rewiring probability $\rho$ for a fixed coupling strength $\epsilon$. For spiking neuron networks, $F$ slightly increased for $\rho < 0.6$ and then reached a plateau. For the other neuron networks, $F$ attained similar values for all investigated rewiring probabilities, except for $\rho<0.1$.

\begin{figure}[!h] \centering
        \includegraphics[ width = 0.9 \columnwidth ]{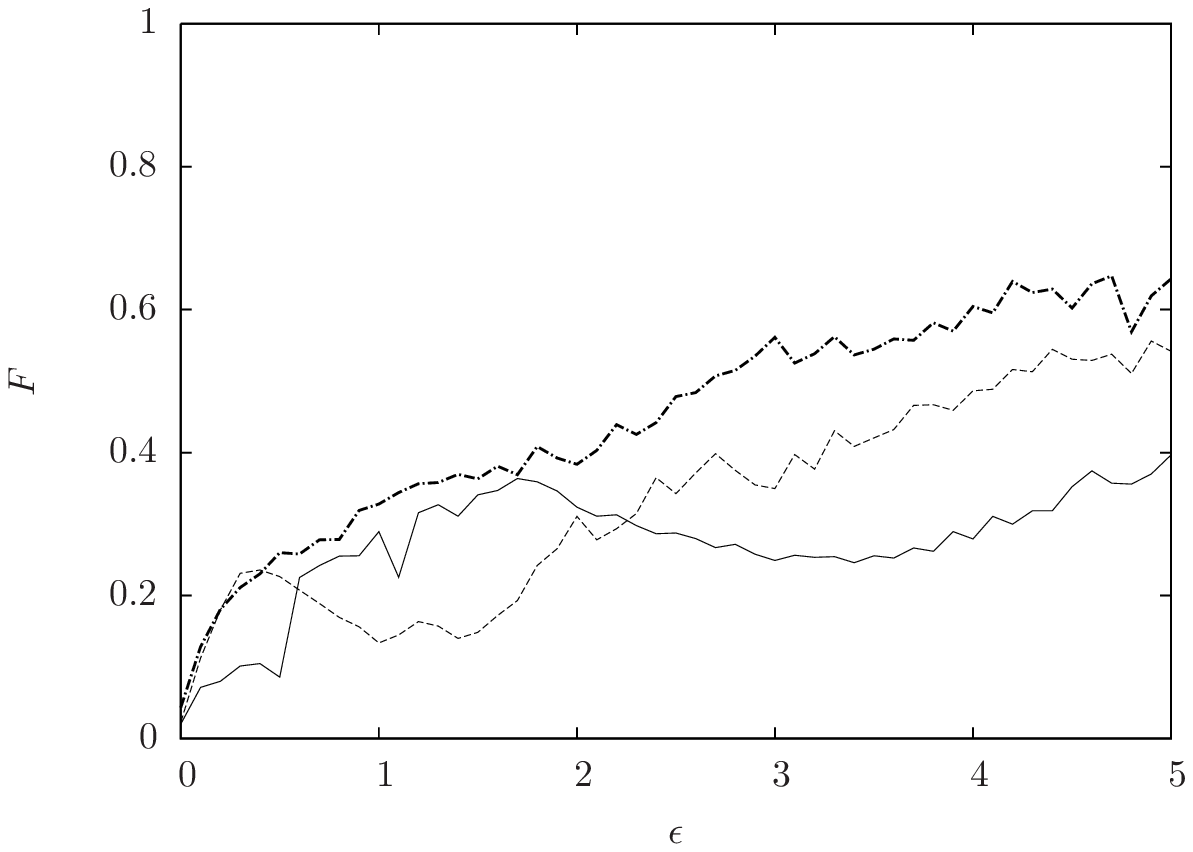}
	\caption{Synchrony $F$ in dependence on the coupling strength $\epsilon$; fixed rewiring probability $\rho = 0.3$. Solid line: spiking neuron network; darker dashed line: bursting neuron network; lighter dashed line: chattering neuron network.}
        \label{cg-hom_cut}
\end{figure} 

In Figure \ref{cg-hom_cut} we show the dependence of network synchrony $F$ on the coupling strength $\epsilon$ for a fixed rewiring probability $\rho$. For networks of chattering neurons, $F$ increased monotonously with increasing $\epsilon$. For small coupling strength ($\epsilon \leq 0.3$), $F$  attained similar values for networks of bursting and chattering neurons. However, for $\epsilon \geq 0.3$, $F$ dropped for networks of bursting neurons even below the levels observed for networks of spiking neurons. For larger values of $\epsilon$ ( $\epsilon \leq 1.5$)  $F$ increased again and approached the values for networks of chattering neurons. 

For networks consisting of spiking neurons, we observed $F$ to increase with increasing $\epsilon$ similarly as observed for chattering neuron networks.  At $\epsilon\approx 1.7$ synchrony of spiking neuron networks decreased since these neurons sometimes showed bursting behavior due to the large input currents generated by firing neurons. For even larger values of $\epsilon$ synchrony $F$ increased again.

\subsection{Inhomogeneous Networks}
\label{cg-Inhomo}

\begin{figure}[!h] \centering
	\includegraphics[ width = 0.9 \columnwidth ]{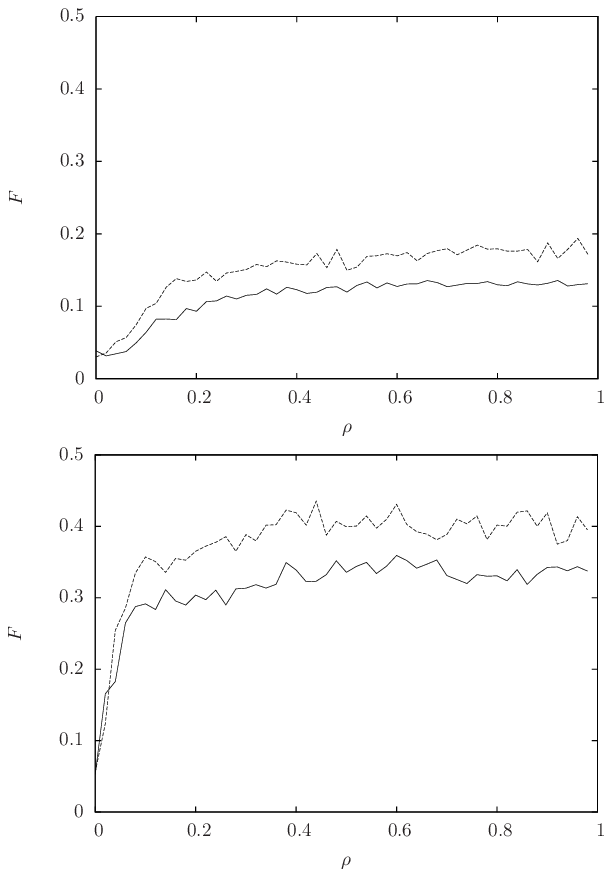}
	\caption{Synchrony $F$ in dependence on the rewiring probability $\rho$ for networks consisting of 70\% spiking and 30\% of bursting/chattering neurons. Solid line: spiking/bursting neuron network; dashed line: spiking/chattering neuron network. Top: $\epsilon=1$ ; bottom: $\epsilon=4$.}
	\label{cg-mixedCutrho}
\end{figure} 

\begin{figure}[!h] \centering
	\includegraphics[ width = 0.9 \columnwidth ]{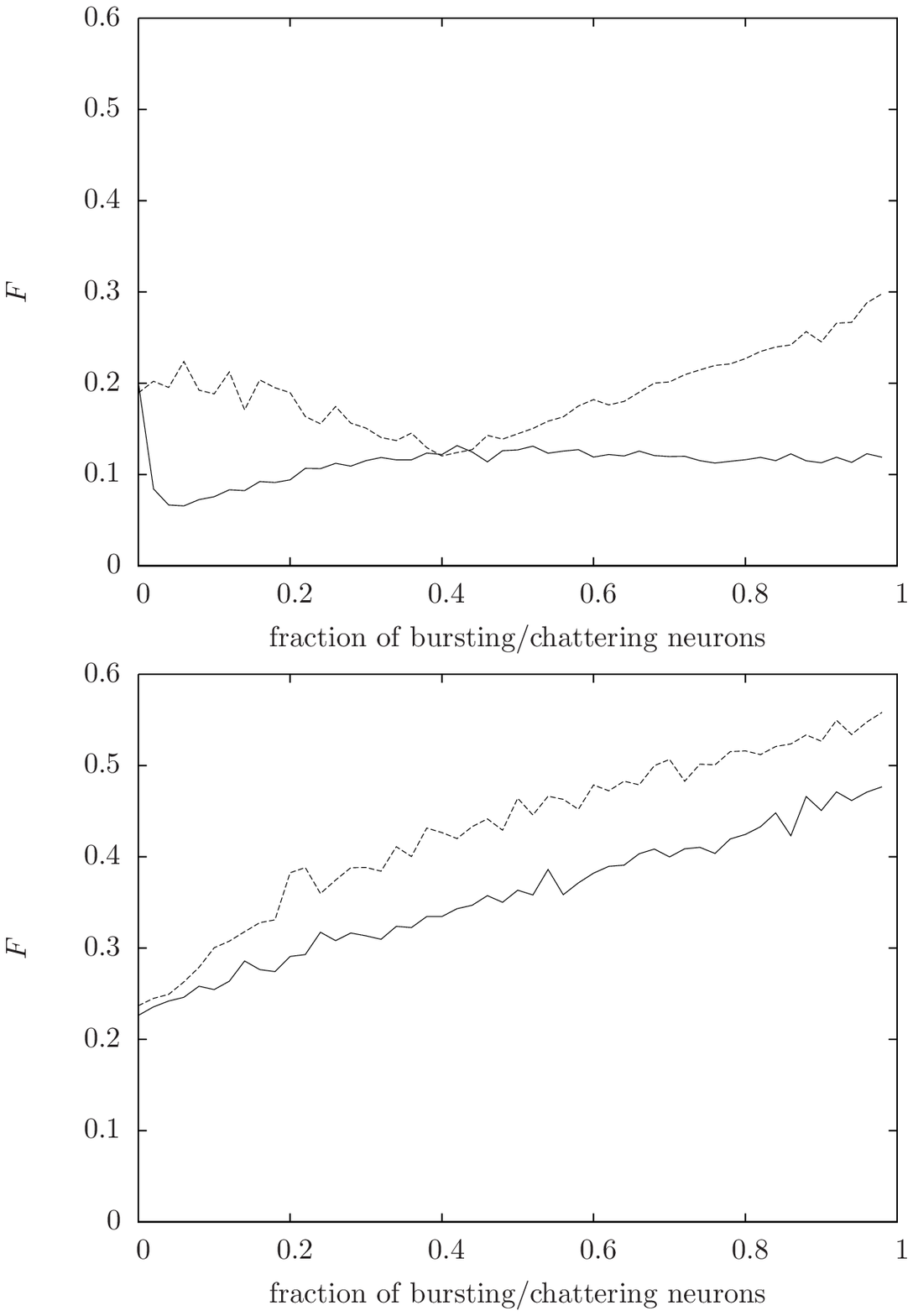}
	\caption{Synchrony $F$ for inhomogeneous networks with $\rho=0.3$ in dependence on the fraction of spiking and bursting/chattering neurons. Solid line: spiking-bursting neuron network; dashed line: spiking-chattering neuron network. Top: $\epsilon=1$; bottom: $\epsilon=4$.}
	\label{cg-mixedCutFrac}
\end{figure}

In inhomogeneous networks we observed synchrony $F$ to increase with an increasing rewiring probability $\rho$ until $\rho \approx 0.2$ and then to approximately stay constant, irrespective of the coupling strength $\epsilon$ (see Figure \ref{cg-mixedCutrho}). Similarly as observed for homogeneous networks, synchrony $F$ attained larger values for larger coupling strength $\epsilon$. We also observed for $\rho>0$ higher values of $F$ for networks containing chattering neurons than for networks containing bursting neurons, which is analogous to our finding for homogeneous networks. 

For a fixed rewiring probability $\rho=0.3$ and large coupling strength ($\epsilon=4$) we observed an approximately linear relationship between synchrony $F$ and the fraction of bursting or chattering neurons in a network (see Figure \ref{cg-mixedCutFrac} bottom). For a small coupling strength $\epsilon=1$, we observed synchrony $F$ first to decrease with the fraction of bursting or chattering neurons and than to increase again(see Figure \ref{cg-mixedCutFrac} top). 

For the spiking-chattering neuron network $F$ decreased until it reached a minimum at around 40\% chattering neurons and then increased to larger values of $F$ than for homogeneous spiking neuron networks. This was, however, not the case for spiking-bursting neuron networks. Here synchrony $F$ decreased with the fraction of bursting neurons until it reached a minimum at about 5\% bursting neurons and then increased but remained smaller than for homogeneous spiking neuron networks. This is in agreement with our findings for homogeneous networks as synchrony was larger for spiking neuron than for bursting neuron networks for coupling strength $\epsilon\in[0.75,2.0]$. 

\section{Conclusion}
We studied complex neuron networks with homogeneous and inhomogeneous local dynamics.
We observed that chattering neuron networks exhibited higher levels of synchrony than bursting neuron networks which in turn exhibited higher levels of synchrony than spiking neuron networks.
In addition, synchrony was higher for small-world and random network configurations than for lattice-like structures.
In inhomogeneous networks, composed of both spiking and bursting neurons, we observed that synchrony may be decreased due to the influence of the bursting neurons.
These observations support the notion, that under certain conditions inhomogeneity, such as in the degree (connectivity) distribution \cite{Motter2005}, in the coupling architecture \cite{Denker2004,Dyhrfjeld-Johnsen2007}, or -- as shown here -- with different node dynamics, hinders synchrony.

\bibliographystyle{ws-procs9x6}

\end{document}